\documentclass[%
reprint,
 amsmath,amssymb,
 aps,
]{revtex4-1}

\usepackage{graphicx}
\usepackage{dcolumn}
\usepackage{bm}

\def \kt {\tilde{k}}
\def \lt {\tilde{l}}

 \usepackage{microtype}

\usepackage{hyperref}
\hypersetup{
	hyperindex,
	breaklinks,
	colorlinks=true,
	linkcolor=blue,
	citecolor=blue,
	bookmarks=true,
	bookmarksopen=true,
	bookmarksopenlevel=2,
	pdfstartpage={1},
	pdfstartview={FitH},
	pdfview={FitH 0},
	pdfauthor={Nikos A. Bakas and Petros J. Ioannou},
	pdftitle={Emergence of large scale structure\\in planetary turbulence},
 }

\begin{document}

\preprint{APS/123-QED}

\title{Emergence of large scale structure\\in planetary turbulence}

\author{Nikolaos A. Bakas}
 \email{nikos.bakas@gmail.com}
\author{Petros J. Ioannou}%
 \email{pjioannou@phys.uoa.gr}
\affiliation{%
 Department of Physics\\
 National and Kapodistrian University of Athens\\
 Panepistimiopolis, Zografos \\
 Athens 15784, Greece
}%

%
%

\date{\today}

\begin{abstract}
Planetary and magnetohydrodynamic drift-wave turbulence is observed to self-organize into large scale
structures such as zonal jets and coherent vortices. In this Letter we present a non-equilibrium statistical
theory, the Stochastic Structural Stability theory (SSST), that can make predictions for the formation and
finite amplitude equilibration of non-zonal and zonal structures (lattice and stripe patterns) in
homogeneous turbulence. This theory reveals that the emergence of large scale structure
is the result of an instability of the interaction between the coherent flow and the associated turbulent field.
Comparison of the theory with nonlinear simulations of a barotropic flow in a $\beta$-plane channel with
turbulence sustained by isotropic random stirring, demonstrates that SSST predicts the threshold parameters
at which the coherent structures emerge as well as the characteristics of the emerging structures (scale,
amplitude, phase speed). It is shown that non-zonal structures (lattice states or zonons) emerge at lower
energy input rates of the stirring compared to zonal flows (stripe states)  and their   emergence affects the dynamics of jet
formation.
\begin{description}
\item[PACS numbers] 47.27.eb, 47.20.Ky, 47.27.De,  52.35.Mw, 89.75.Fb, 92.60.Bh, 92.10.A-
\end{description}
\end{abstract}

\pacs{Valid PACS appear here}
\keywords{beta plane turbulence, jets, non-zonal coherent structures}
\maketitle


Turbulence in planetary atmospheres and in plasma flows is commonly observed to
be organized into large scale unidirectional (zonal) jets with long-lasting coherent eddies or vortices embedded in
them~\citep{Vasavada-05,Diamond-05}.
The jets control the transports of heat and chemical
species in planetary atmospheres and separate the high temperature plasma from the cold containment vessel
wall in magnetic plasma confinement devices. 
It is therefore important to understand the mechanisms for the emergence, equilibration and maintenance
of these coherent structures. In this Letter we present a theory that  predicts
the regime changes occurring in the turbulent flow as well as the amplitude, structure and propagation characteristics
of both the zonal jets and the non-zonal coherent structures that form in the flow. We then test this theory
against non-linear simulations in a simple model of forced planetary and plasma turbulence.

The simplest model that captures the  turbulent dynamics  and its  interaction with the zonal jets and the coherent structures, is
the stochastically forced barotropic vorticity equation on a plane tangent to the surface of a rotating planet:
\begin{equation}
\partial_t\zeta+\psi_x\zeta_y-\psi_y\zeta_x+\beta\psi_x=-r\zeta-\nu\Delta^2 \zeta+f.\label{eq:derivation1}
\end{equation}
The relative vorticity is $\zeta = \Delta\psi$, $\psi$ is  the streamfunction, $\Delta=\partial_{xx}^2+
\partial_{yy}^2$ is the horizontal Laplacian,  $x$ is in the zonal (east-west) direction and $y$ is in the meridional
(north-south) direction,  $\beta = 2 \Omega \cos \phi_0/ R$ is
the gradient of planetary vorticity,  $\Omega$ the rotation rate of the planet, $\phi_0$ the latitude of the $\beta$-plane
and $R$ is the radius of the planet. Equation (\ref{eq:derivation1}) governs the dynamics of non-divergent motions at
the midlatitudes of the planet and is  also the infinite effective Larmor radius limit of the Charney-Hasegawa-Mima equation
that governs  drift-wave turbulence in plasmas. We are assuming linear damping with coefficient,
$r$, representing the
Ekman drag induced by
the horizontal boundaries and hyper-diffusion with coefficient $\nu$ that dissipates the energy flowing into
unresolved scales.
The forcing term $f$ is necessary to sustain turbulence, and may parameterize processes that have not been
included in the dynamics, such as the vorticity forcing from small scale convection. In many previous studies, this exogenous
excitation is taken as a temporally delta correlated and spatially homogeneous and isotropic random stirring. We will
follow the same forcing protocol in this Letter and consider an  isotropic ring forcing that is injecting energy at rate
$\epsilon$ in a narrow ring of wavenumbers of width $\Delta K_f$ around the total wavenumber $K_f$.

\begin{figure}
\hspace*{-5mm}\includegraphics[width=3.2in]{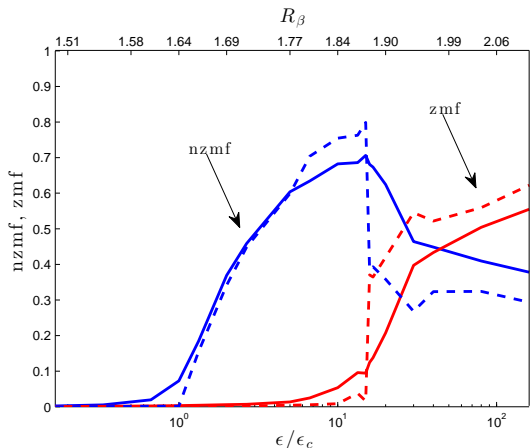}
\vspace*{-5mm}\caption{ The \mbox{zmf} (red lines) and \mbox{nzmf} (blue lines) indices  as a function of energy
input rate $\epsilon/\epsilon_c$ and the corresponding zonostrophy parameter $R_\beta$
for the non-linear (solid lines) and SSST (dashed lines) integrations. The critical value
$\epsilon_c$ (corresponding to $R_\beta=1.64$)
is the energy input rate at which the SSST predicts structural instability of the homogeneous turbulent state. Zonal jets
emerge here for $\epsilon > \epsilon_{nl}$, with $\epsilon_{nl} =15.9 \epsilon_c$ (corresponding to $R_\beta=1.88$). }
\label{fig:zmf}
\end{figure}

\begin{figure}
\includegraphics[width=3.4in]{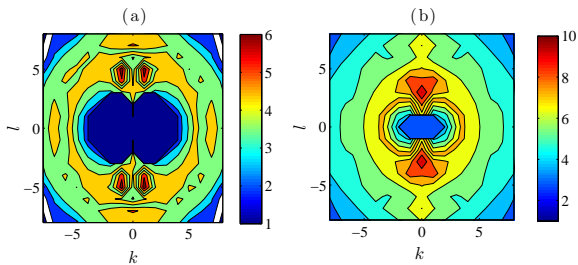}
\vspace*{-5mm}\caption{ Time averaged energy power spectra, $\log_{10}(\hat{E}(k,l))$, obtained from  non-linear
simulation of Eq.~\eqref{eq:derivation1} at  (a) $\epsilon/\epsilon_c=2.6$ and (b) $\epsilon/\epsilon_c=30$. In (a)
the flow is dominated by a $(|k|,|l|)=(1,5)$ non-zonal coherent structure. In (b) the flow is dominated by  a coherent zonal flow
at $(k,|l|)=(0,3)$.}
\label{fig:spectra}
\end{figure}
We solve \eqref{eq:derivation1} in a doubly periodic domain of size $2\pi\times 2\pi$. The calculations presented
in this Letter are for $\beta=10$, $r=0.01$, $\nu=2\cdot 10^{-6}$,
$K_f=8$ and $\Delta K_f=1$, which are reasonable planetary parameters. The results discussed were verified to be insensitive to
the forcing protocol. To illustrate some of the
characteristics of the turbulent flow and the emergence
of coherent structures, we consider two indices. The first is the zonal mean flow index~\citep{Srinivasan-Young-12} defined as
the ratio of the energy of zonal jets over the total energy,
$\mbox{zmf}=\frac{\sum_l \hat{E}(k=0,l)}{\sum_{kl}\hat{E}(k,l)}$, where $\hat{E}$ is the time averaged energy power spectrum
of the flow and $k$, $l$ are the zonal and meridional wavenumbers
respectively. The second is the non-zonal mean flow index defined as the ratio of the energy of the non-zonal modes with
scales lower than the scale of the forcing over the total energy: $\mbox{nzmf}=\frac{\left(\sum_{kl: K<K_f}\hat{E}(k,l)-\sum_l \hat{E}(k=0,l)\right)}{\sum_{kl}\hat{E}(k,l)}.$
Figure~\ref{fig:zmf} shows both indices as a function of the energy input rate  $\epsilon$ and the corresponding value of the
non-dimensional zonostrophy parameter $R_{\beta} = 0.7 (\epsilon \beta^2/ r^5)^{1/20}$, which has been used in previous studies
to characterize the emergence and structure of zonal jets in planetary turbulence \citep{Galperin-etal-10,Scott-Dritchel-2012}.
For $\epsilon$ smaller than a critical value $\epsilon_c$ (corresponding to $R_\beta=1.64$), the turbulent flow is homogeneous and remains
translationally invariant in both directions. When $\epsilon>\epsilon_c$, the translational symmetry of the
flow is broken and non-zonal structures form with scales larger than the scale of the forcing, as indicated by the rapid increase
of the nzmf index.
\begin{figure}[t]
\hspace*{-5mm}\includegraphics[width=3.8in]{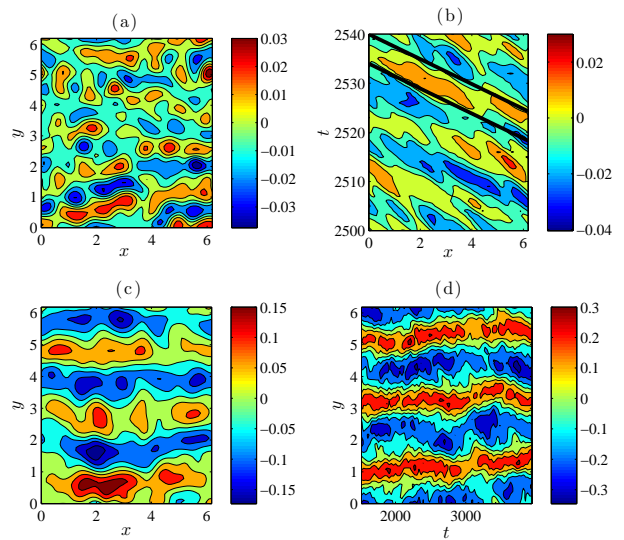}
\caption{ (a) Snapshot of the streamfunction $\psi(x,y,t)$ and (b) Hovm\"oller diagram of
$\psi(x,y=\pi/4,t)$ obtained from  non-linear simulation for $\epsilon/ \epsilon_c=2.6$. The thick lines
in (b) correspond to the phase speed obtained from the stability equation (\ref{eq:dispersion}). (c) Snapshot of the streamfunction
$\psi(x,y,t)$ and (d) Hovm\"oller diagram of
the $x$-averaged $\overline{\psi(y,t)}$, which is  obtained from non-linear simulations at $\epsilon/ \epsilon_c=30$. }
\label{fig:NL_snap1}
\end{figure}
The time averaged power
spectrum, shown in Fig.~\ref{fig:spectra}(a) for $\epsilon=2.6\epsilon_c$, has pronounced  peaks at
$(|k|, |l|)=(1, 5)$ that correspond to coherent structures propagating westward (cf.  Fig.~\ref{fig:NL_snap1}(a),(b))
with approximately the Rossby wave phase speed for this wave.
However, at larger $\epsilon$  the propagation speed of these structures departs from that of Rossby waves.
The presence and properties of such non-linear waves in similar simulations, were also reported recently~\citep{Sukoriansky-etal-2008}.
For $\epsilon>\epsilon_{nl}$ indicated in Fig. \ref{fig:zmf} (corresponding to $R_\beta=1.88$), robust zonal jets emerge
as is shown by the rapid increase of the zmf index. For example
the peaks at $(k, |l|)=(0, 3)$ in the spectrum  of  Fig.~\ref{fig:spectra}(b)  correspond to coherent zonal jets
(cf.  Fig.~\ref{fig:NL_snap1}(c),(d)). From Fig.~\ref{fig:zmf} we see that while the jets contain over half of the
total energy, substantial power remains in non-zonal structures. Previous studies, refer to
the coherent  non-zonal structures in this regime in which strong zonal jets dominate the flow (referred to as
zonostrophic regime \cite{Galperin-etal-06})  as satellite modes~\cite{Danilov-04} or zonons
\citep{Sukoriansky-etal-2008,Galperin-etal-10}.

The emergence of jets has been described in previous studies in terms of an anisotropic inverse energy
cascade~\citep{Rhines-75,Vallis-93,Nazarenko-09},
or in terms of inhomogeneous mixing of vorticity~\citep{Dritchel-2008,Scott-Dritchel-2012}, or in terms of a direct
transfer of energy from small scale waves into the zonal jets, through either non-linear interactions between finite
amplitude Rossby waves~\citep{Gill-74,Connaughton-09}, or through shear straining of the small scale waves by the
jet~\citep{Robinson-06}. However, the mechanism for the emergence of the non-zonal structures remains elusive.
Statistical equilibrium theory applied in the absence of forcing and dissipation, has been able to predict both jets
and coherent vortices as maximum entropy structures \citep{Bouchet-2012} and a recent study has shown correspondence
of the theoretical results with non-linear simulations in the limit of weak forcing and dissipation \citep{Bouchet-Simonnet-09}.
Nevertheless, the relevance of these results in planetary and plasma flows that are strongly forced and dissipated and are
therefore out of equilibrium remains to be shown. In this Letter we present results of an alternative, non-equilibrium
statistical theory, that is termed as Stochastic Structural Stability Theory (SSST) \citep{FI-03,FI-07} or Second Order
Cumulant Expansion theory (CE2) \citep{Marston-etal-2008,Srinivasan-Young-12, Marston-2012}. While recent studies have
demonstrated that SSST can predict the structure of zonal flows in turbulent fluids
\citep{Srinivasan-Young-12,Constantinou-etal-2012,Tobias-etal-2012}, the results
presented in this Letter, demonstrate that an extended version of SSST can predict the emergence of both zonal and non-zonal
coherent structures in planetary turbulence and can capture their finite amplitude manifestations. The emergence of non-zonal and zonal structures described above is similar to formation of the lattice and
stripe patterns  in homogeneous thermal non-equilibrium systems \citep{Cross-Greenside-2009}.  The analogy
between the  formation of stripes  and zonal jets  has been recently emphasized using  SSST dynamics  \citep{Parker-Krommes-2013}.
In this letter we formulate the SSST dynamics that can produce lattice states in fully turbulent flows.

SSST describes the statistical dynamics of  the first two equal time cumulants of Eq.~(\ref{eq:derivation1}). The first
cumulant  is $Z(\mathbf{x},t) \equiv\left< \zeta\right>$ (the brackets denote an ensemble average) and the second cumulant
$C(\mathbf{x}_1, \mathbf{x}_2, t)\equiv \left<\zeta_1'\zeta_2'\right>$  is a function of the vorticity deviation
$\zeta_i'=\zeta_i-Z_i$ at the two points $\mathbf{x}_i=(x_i, y_i)$ ($i=1,2$) at time $t$. It can be shown  from \eqref{eq:derivation1}  that
the equations for the evolution of the two cumulants are:
\begin{align}
\partial_t Z&+UZ_x+V(\beta+Z_y)+rZ+\nu\Delta^2Z=\nonumber\\
&\;= \partial_x\left(\partial_{y_1}\Delta_2^{-1}C\right)_{\mathbf{x}_1=\mathbf{x}_2}-\partial_y\left(\partial_{x_1}\Delta_2^{-1}C\right)_{\mathbf{x}_1=\mathbf{x}_2},\label{eq:Q_evo2}\\
\partial_t C&=(A_1+A_2)C+\Xi.\label{eq:cov_evo2}
\end{align}
The linear operator
\begin{align}
A_i=-U_i\partial_{x_i}-V_i\partial_{y_i}-&(\beta+Z_{y_i})\partial_{x_i}\Delta_i^{-1}+\nonumber\\
&+
Z_{x_i} \partial_{y_i}\Delta_i^{-1}-r-\nu\Delta_i^2, \label{eq:op_A}
\end{align}
acts at the points $\mathbf{x}_i=(x_i, y_i)$ and governs the dynamics of linear perturbations about the instantaneous
mean flow $\mathbf{U}=[U, V]=[-\partial_y\left< \psi \right >, \partial_x \left <\psi\right>]$. In (\ref{eq:cov_evo2}), $\Xi$
contains the covariance
of the external forcing  and terms related to third order cumulants. A second order closure is obtained if the third order
cumulant is  ignored and   $\Xi$ is set to be the spatial covariance of the stochastic forcing $f$.
In most earlier studies of SSST or CE2, the ensemble average was assumed to represent a zonal average. In this Letter, we
adopt the more general interpretation that the ensemble average represents a Reynolds average with
the ensemble mean representing coarse-graining. This interpretation has been adopted in the SSST study of turbulence in
baroclinic flows \citep{Bernstein-2009,Bernstein-Farrell-2010}. With this interpretation of the ensemble mean, the SSST system
(\ref{eq:Q_evo2})-(\ref{eq:cov_evo2}) provides  the statistical dynamics of the interaction of the ensemble average field,
which can be a zonal or a non-zonal coherent structure, with the fine-grained field, represented in the theory through its
covariance $C$. The SSST system defines an autonomous dynamics and its fixed points define a new type of turbulent statistical
equilibria. While these equilibria formally exist only in the infinite ensemble limit, it has been shown that their
characteristics manifest even in single realizations of the turbulent system.
The structural stability
of these turbulent equilibria can be addressed in SSST by studying their stability. Specifically, when an
equilibrium of the SSST equations becomes unstable, the turbulent flow bifurcates to a different attractor. This theory therefore
predicts parameters of the physical system which can lead to abrupt reorganization of  the turbulent flow.

The SSST equations (\ref{eq:Q_evo2})-(\ref{eq:cov_evo2}), admit for $\nu=0$ the  simple equilibrium
\begin{equation}
U^E=V^E=0,~~C^E={\Xi\over 2r},\label{eq:equil}
\end{equation}
that has zero large scale flow and  a homogeneous eddy field with spatial covariance
dictated directly from the  forcing. We now investigate the SSST stability of this equilibrium as a function of the energy
input rate, $\epsilon$, and relate the outcome of this stability analysis to the results in the non-linear simulations of
(\ref{eq:derivation1}). The stability of the homogeneous equilibrium (\ref{eq:equil}) is assessed by introducing small
perturbations of the form $[\delta Z, \delta C]=[\delta Z_{nm}, \delta C_{nm}]e^{in(x_1+x_2)/2+im(y_1+y_2)/2}e^{\sigma t}$
to the SSST equations (\ref{eq:Q_evo2})-(\ref{eq:cov_evo2}) linearized about the equilibrium (\ref{eq:equil}) and calculating
the eigenvalue $\sigma$. When $\Re(\sigma)>0$, the structure with $(x, y)$ wavenumbers $(n, m)$ is unstable and
will emerge. It can be shown that  $\sigma$ satisfies the non-dimensional equation
\begin{widetext}
\begin{equation}
\frac{\tilde\epsilon K_f}{2 \pi \Delta K_f} \sum_{k,l}\frac{(\tilde m \kt-\tilde n \tilde l)\left[\tilde n\tilde m(\tilde k_+^2-\tilde l_+^2)+(\tilde m^2-\tilde n^2)\tilde k_+\lt_+
\right](1-\tilde N^2/\tilde K^2)}{2i\kt_+(\kt_+\tilde n+\lt_+\tilde m)-i\tilde n\left(\tilde{K}^2+\tilde{K}_s^2\right)/2+(\tilde\sigma+2)\tilde{K}^2 \tilde{K}_s^2} =
(\tilde\sigma+1)\tilde N^2-i\tilde n.\label{eq:dispersion}
\end{equation}
\end{widetext}
In this equation $(\tilde n, \tilde m,\kt,\lt)=(n, m,k,l)r/ \beta$, $\tilde\sigma=\sigma /r$ are the non-dimensional wavenumbers 
and growth rate respectively, $\tilde K^2=\kt^2+\lt^2$,
$\tilde{K}_s^2=(\kt+\tilde n)^2+(\lt+\tilde m)^2$, $\tilde N^2=\tilde n^2+\tilde m^2$, $\kt_+=\kt+\tilde n/2$ and $\lt_+=\lt+\tilde m/2$ and 
the summation is over integer values of  $(k, l)$
satisfying $|\tilde K-(K_fr/\beta)|<\Delta K_fr/\beta$ \footnote{Hyperdiffusion can be readily included in
\eqref{eq:dispersion} in order to obtain correspondence with the  nonlinear simulations}.
The non-dimensional energy input rate $\tilde\epsilon=\epsilon\beta^2/r^5$, which  is the bifurcating parameter in 
this Letter, is related to the zonostrophy parameter through $R_\beta=0.7\tilde \epsilon^{1/20}$. For $n=0$ the stability 
equation \eqref{eq:dispersion} reduces to the equation that determines  the emergence of zonal flows 
\citep{Bakas-Ioannou-2011,Srinivasan-Young-12}.%


\begin{figure}
\hspace*{-5mm}\includegraphics[width=3.6in]{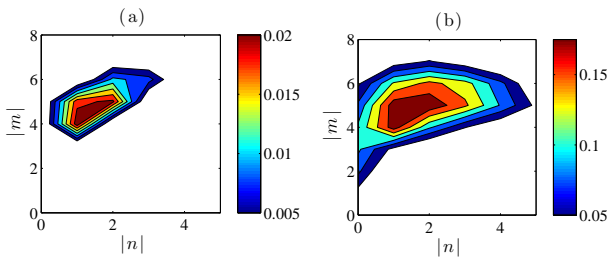}
\vspace*{-8mm}\caption{The growth rate, $\Re(\sigma)$ as a function of the integer valued wavenumbers $|n|$, $|m|$ of the coherent structures for (a)
$\epsilon/ \epsilon_c=2.6$ and (b) $\epsilon/ \epsilon_c=30$  (only positive values of $\Re(\sigma)$ are shown). Unstable waves with
$|n|>0$ have $\Im(\sigma)>0$ corresponding to retrograde propagation.}
\label{fig:growth}
\end{figure}
%
%

\begin{figure}[t]
\includegraphics[width=3.6in]{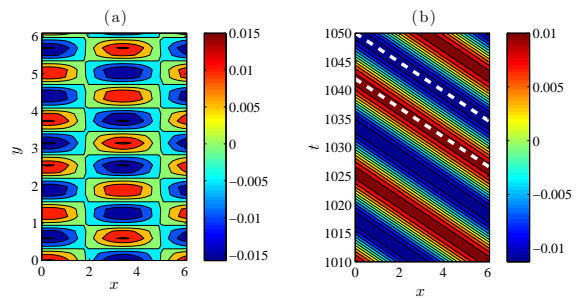}
\vspace*{-5mm}\caption{(a) Snapshot of the streamfunction $\psi(x,y,t)$ and (b) Hovm\"oller diagram of
$\psi(x,y=\pi/4,t)$ obtained from the SSST integrations for $\epsilon/ \epsilon_c=2.6$. The thick dashed lines
in (b) show the phase speed obtained from the stability equation \eqref{eq:dispersion}.}
\label{fig:PL_snap1}
\end{figure}

For small values of the energy input rate $\tilde\epsilon$,  $\Re(\sigma)<0$ and the
homogeneous state is stable. At  $\tilde\epsilon_c$ the homogeneous flow becomes
SSST unstable, symmetry breaking occurs, and coherent structures emerge. The critical  $\tilde\epsilon_c$ is defined as  $\mbox{min}_{(n, m)} \tilde\epsilon_t$, where $\tilde\epsilon_t$
is the  energy input rate that renders wavenumbers $(n, m)$ neutral ($\Re(\sigma)=0$).
The critical $\tilde \epsilon_c$  depends in general on the forcing characteristics
and for the ring forcing at $K_f=8$, $\tilde\epsilon_c=2.48\cdot 10^7$ or $R_\beta=1.64$ \footnote{
It can be shown that  $\tilde \epsilon_c$ is a rapidly decreasing function of $K_f$ and  the critical value for instability asymptotically  occurs
 at  $\lim_{K_f \rightarrow \infty} \tilde \epsilon_c = 23$ or  $R_\beta=0.82$.}.
The growth rates as a function of the
integer valued wavenumbers, $(n, m)$, of the structure are shown in Fig.~\ref{fig:growth}. For $\epsilon/\epsilon_c=2.6$, the structure with
the largest growth rate, is non-zonal with $(|n|,|m|)=(1, 5)$ and has $\Im({\sigma})=0.4$, implying retrograde propagation of
the eigenstructure. Note also that for this energy input rate, the zonal flows are SSST stable and jets are  not expected to form.
For $\epsilon/\epsilon_c=30$, both zonal jets and non-zonal structures are unstable, but the zonal jets have smaller growth
rates compared to the non-zonal structures \footnote{This  is always  true for \beta >\beta_{min}$.
For the given isotropic
forcing, $\beta_{min}=4.5 r K_f $.}. The zonal jets  are stationary ($\Im(\sigma)=0$), in contrast to the
non-zonal coherent structures that always propagate in the retrograde direction. Numerical integration of the SSST system
(\ref{eq:Q_evo2})-(\ref{eq:cov_evo2}), shows that for $\epsilon>\epsilon_c$ the unstable structures typically
equilibrate at finite amplitude after an initial period of exponential growth. As a result (\ref{eq:Q_evo2})-(\ref{eq:cov_evo2})
admit in general multiple equilibria. Figure \ref{fig:PL_snap1}(a) shows the equilibrium structure with the
largest domain of attraction, when $\epsilon/ \epsilon_c=2.6$. This structure coincides in this case with the finite amplitude
equilibrium of the fastest growing $(|n|, |m|)=(1, 5)$ eigenfunction and propagates as illustrated in Fig.~\ref{fig:PL_snap1}(b)
in the retrograde direction with a speed approximately equal to the phase speed of this unstable eigenstructure. A proxy for the
amplitude of these equilibrated structures are the $\mbox{zmf}$ and $\mbox{nzmf}$ indices calculated for
the SSST integrations that are shown in Fig.~\ref{fig:zmf}. As the energy input rate increases, the non-zonal structures
equilibrate at larger amplitudes. However, for $\epsilon>\epsilon_{nl}$, the equilibria with the largest domain of attraction
are zonal jets and the flow is dominated by these structures (cf. Fig.~\ref{fig:zmf}).

The results of the SSST analysis are now compared to  non-linear simulations. First of all, the SSST stability analysis
accurately predicts the critical $\epsilon_c$  for the emergence of non-zonal structures in the non-linear simulations as shown in
Fig.~\ref{fig:zmf}. The finite amplitude equilibria obtained when $\epsilon>\epsilon_c$ also correspond to the dominant structures
in the non-linear simulations. For $\epsilon/ \epsilon_c=2.6$, the spectra in the nonlinear simulations show significant power
at $(|n|, |m|)=(1, 5)$, which is the SSST structure  with the largest domain of attraction. Remarkably, the phase speed of
the $(1, 5)$ waves observed in the non-linear simulations and the amplitude of these
structures as illustrated by the $\mbox{nzmf}$ index are approximately equal to the phase speed and amplitude of the corresponding
SSST equilibrium structure (cf.  Figs.~\ref{fig:zmf}, \ref{fig:NL_snap1} and \ref{fig:PL_snap1}). For $\epsilon > \epsilon_{nl}$, in both nonlinear and
SSST simulations  zonal jets emerge and the power of the non-zonal structures is substantially reduced.
Comparison of the number of jets and their amplitude between the SSST and the nonlinear simulations also shows good
agreement (not shown). This demonstrates that the SSST system can predict the amplitude and characteristics of both the non-zonal
and the zonal structures that emerge in the turbulent flow.

While the regime transition that occurs at $\epsilon_c$ is predicted by the stability equation (\ref{eq:dispersion}),
the second transition, which is associated with the emergence of zonal flows and occurs at $\epsilon_{nl}$,
is more intriguing. The stability equation (\ref{eq:dispersion}) predicts that the zonal structures become unstable at
$\epsilon_{sz}=4\epsilon_c<\epsilon_{nl}$. In previous studies of SSST dynamics restricted to the
interaction between zonal flows and turbulence, these initially unstable structures were found to equilibrate at finite
amplitude \cite{Srinivasan-Young-12,Constantinou-etal-2012}. Preliminary calculations show that within the context
of this generalized SSST analysis that takes into account the dynamics of non-zonal structures as well, these equilibria
are found to be saddles that are stable to zonal but unstable to non-zonal perturbations. The threshold for the emergence of
jets in the SSST integrations and in the nonlinear simulations is therefore determined as the energy input rate at which an
SSST stable, finite amplitude zonal jet equilibrium exists. A method to correctly obtain the critical input rate
$\epsilon_{nl}$ has been recently developed \cite{Constantinou-etal-2012}. It starts by recognizing that for
$\epsilon_c<\epsilon<\epsilon_{sz}$, the spectrum in the turbulent flow is modified ($C^E\neq \Xi /2r$) and is given by the
covariance $\tilde{C}^E$ associated with the finite amplitude equilibria similar to the ones shown in Fig.~\ref{fig:PL_snap1}.
If this modification is taken into account, then the stability analysis around the equilibrium $[U^E=0, \tilde{C}^E]$
correctly predicts $\epsilon_{nl}$.

In summary, we presented a theory that shows that large scale structure in barotropic planetary and drift-wave turbulence
arises  through systematic self-organization of the turbulent Reynolds stresses, through  non-local interactions
and in the absence of cascades.   The theory allowed the determination of conditions for
the emergence of non-zonal coherent structures in homogeneously forced flows and we have demonstrated, through comparison with nonlinear simulations,  that
it  predicts both the emergence  and the finite amplitude equilibration of  structure. An advance made in this Letter is the development of the theoretical framework
that accounts for the emergence of non-zonal states in homogeneous turbulence. These non-zonal (or lattice) states were found to propagate
westward, and  their relation  to westward propagating vortex rings in the ocean and coherent vortices in planetary atmospheres will be the subject of
future research. The homogeneous turbulent flow was found to be more unstable to non-zonal structure. We think  that the finite amplitude  non-zonal states
are susceptible to secondary instability at higher supercriticality, and as a result the prevalent structures in planetary flows are zonal jets.

\begin{acknowledgments}
This research was supported by the EU FP-7 under the PIRG03-GA-2008-230958 Marie Curie Grant. The authors acknowledge
the hospitality of the Aspen Center for Physics supported by the NSF (under grant No. 1066293), where part of this work was done.
The authors would also like to thank Navid Constantinou and Brian Farrell for fruitful discussions.
\end{acknowledgments}

\end{document}